\documentclass[pre,superscriptaddress]{revtex4}
\usepackage{latexsym}
\usepackage{graphicx}
\begin{document}
\title{Sole-Manrubia model of biological evolution: some new insights}

\author{Debashish Chowdhury}
\affiliation{Department of Physics, Indian Institute of Technology, Kanpur 20801
6, India.}

\author{Dietrich Stauffer}
\affiliation{Institute for Theoretical Physics, University of Cologne, 50923 K\"
oln, Germany.}


\begin{abstract} 
The Sole-Manrubia model of ``macro''-evolution describes the origination, 
evolution and extinction of species on geological time scales. We report 
some properties of this model which provide deep insight into this simple 
model which captures several realistic features.

\end{abstract}


\maketitle

\section{Introduction}

Several fundamental questions on the origination, evolution and 
extinction of species have been addressed in the recent years by 
theoretical treatments of {\it dynamic network} models using the 
concepts and techniques of non-equilibrium statistical mechanics 
\cite{drossel,newman}. Kauffman \cite{kauffman} pioneered a model
which describes biological evolution as a ``walk'' in a ``fitness 
landscape'' and argued that the eco-system adaptively evolves so 
as to be delicately poised ``at the edge of chaos''. On the other 
hand, Bak and Sneppen \cite{bak}, suggested an alternative way of 
implementing the Darwinian principle of selection  (the ``survival 
of the fittest''); this dynamics naturally leads the eco-system to a 
``self-organized'' critical state\cite{jensen}. These models 
implicitly assume a primarily biotic cause of extinctions. However, 
models based on purely abiotic causes, which arise from ``environmental 
stresses'' also lead to similar pattern of extinctions of species 
\cite{newman,newman2}.

A few years ago Sole and Manrubia (SM) \cite{sole} introduced one such
model of ``macro''-evolution that takes into account the inter-species
interactions explicitly. They reported several statistical properties 
of the model as well as their biological implications. In this paper  
we report some other statistical properties of this model; some of 
these features are qualitatively similar to the corresponding results 
obtained in a more recent model \cite{chowstau}. Moreover, we 
re-interpret some of the quantities in their approach to extract some 
relevant biological implications of the model. Furthermore, we introduce 
a new correlation function throwing light on the ``volatility'' of the 
fitness and the level of ``coherence'' in the changes of fitness of the 
species in this model.

\section{The model}

For the sake of completeness, we briefly present the SM model and 
its main published results. 

The system consists of $N$ species, each labelled by an index $i$ 
($i = 1,2,...N$). The states of the $i$-th species is represented 
by a two-state variable $S_i$; $S_i = 0$ or $1$ depending on whether 
it is extinct or alive, respectively. The inter-species interactions 
are captured by the interaction matrix ${\bf J}$; the element 
$J_{ij}$ denotes the influence {\it of} the species $j$ {\it on} 
the species $i$.  If $J_{ij} > 0$ while, simultaneously, $J_{ji} < 0$ 
then $i$ is the predator and $j$ is the prey. On the other hand, if 
both $J_{ij}$ and $J_{ji}$ are positive (negative) they the two 
species cooperate (compete). 

The dynamics of the system consists in updating the states of the 
system (i.e., to determine the state at the time step $t+1$ from a 
complete knowledge of the state at the time $t$) in the following 
three steps:\\

\noindent {\it Step (i)}: One of the input connections $J_{ij}$ for 
each species $i$ is picked up randomly and assigned a new value 
drawn from the uniform distribution in the interval $[-1,1]$, 
irrespective of its previous magnitude and sign.

\noindent {\it Step (ii)}: The new state of each of the species is 
decided by the equation 
\begin{equation}
S_i(t+1) = \Theta \biggl(\sum_{j=1}^N J_{ij} S_j(t) - \theta_i\biggr)
\end{equation}
where $\theta_i$ is a threshold parameter for the species $i$ and 
$\Theta(x)$ is the standard step function, i.e., $\Theta(x) = 1$ 
if $x > 0$ but zero otherwise. If $S(t+1)$ becomes zero for $m$ 
species, then an extinction of size $m$ is said to have taken 
place.

\noindent {\it Step (iii)}: All the niches left vacant by the extinct 
species are refilled by copies of one of the randomly selected 
non-extinct species.

Sole and Manrubia recorded extinctions of sizes as large as $500$ 
and the distributions of the sizes of these extinctions could be 
fitted to a power law of the form $N(m) \sim m^{-\alpha}$ with an 
exponent $\alpha \simeq 2.3$. Moreover the periods of stasis 
$t_s$ were also found to obey a power law $N(t_s) \sim t_s^{-\gamma}$ 
with the exponent $\gamma \simeq 3.0$. However, surprisingly, in 
none of their papers \cite{sole}, Sole and his collaborators 
reported the distributions of the lifetimes of species which, 
according to some claims (see, for example, refs.\cite{drossel,newman} 
for references to the experimental literature and data analysis), also 
follows a power law.

\begin{figure}[tb]
\begin{center}
\includegraphics[angle=-90,width=0.65\columnwidth]{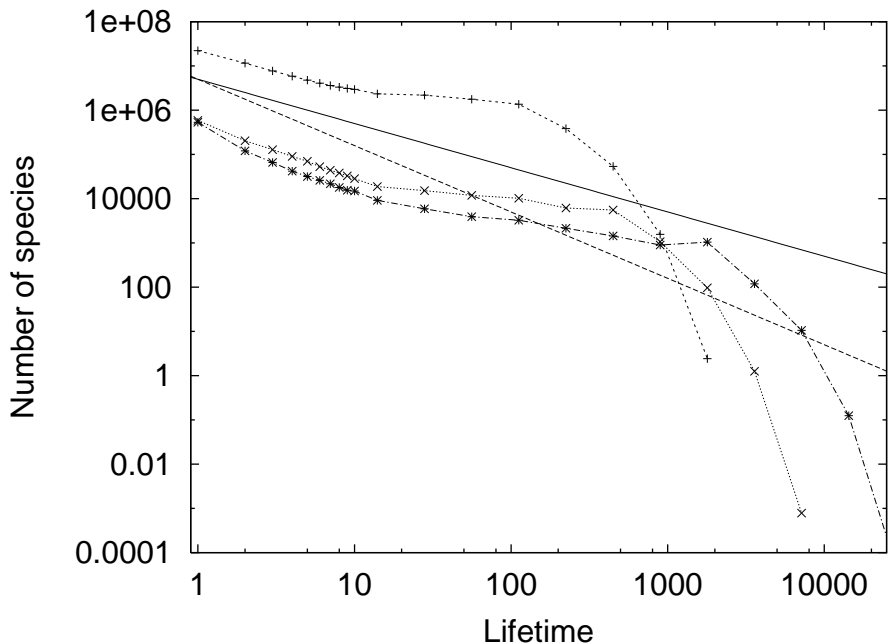}
\end{center}
\caption{Log-log plots of the distributions of the lifetimes of the
species in the SM model with 
$N = 100, ~t_w = 10^6, ~t_m = 8 \times 10^6$ ($+$), 
$N = 300, ~t_w = 10^4, ~t_m = 8 \times 10^4$ ($\times$), 
$N = 1000, ~t_w = 10^4, ~t_m = 8 \times 10^4$ ($\ast$). 
The solid line (with slope $-1$) and the dashed line (with slope 
$-1.5$) are drawn just for the comparison of the simulation data 
with power law of the form $N_t \sim t^{-x}$. }
\label{fig-1}
\end{figure}

\begin{figure}[tb]
\begin{center}
\includegraphics[angle=-90,width=0.65\columnwidth]{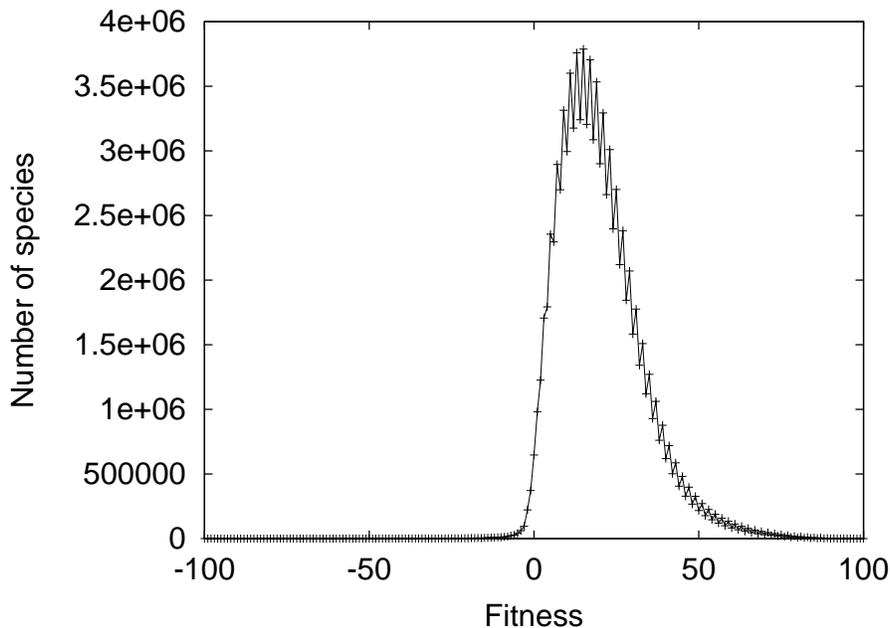}
\end{center}
\caption{Distribution of the fitnesses of the species in the 
SM model with $N = 100, ~t_w = 10^5, ~t_m = 10^6$. 
}
\label{fig-2}
\end{figure}

\begin{figure}[tb]
\begin{center}
\includegraphics[angle=-90,width=0.65\columnwidth]{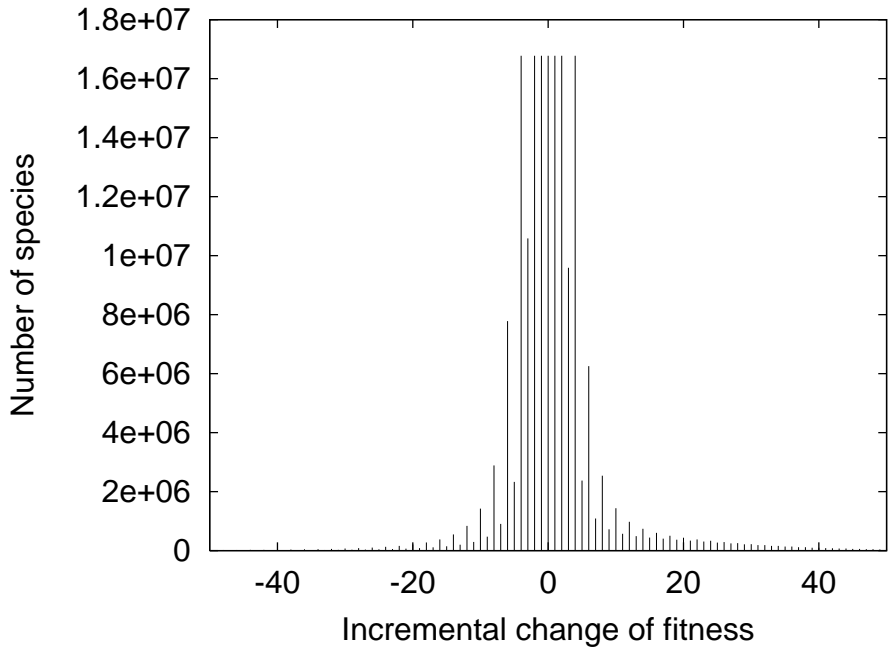}
\end{center}
\caption{Distributions of the incremental changes (per unit time step) 
in the fitness of the species in the 
SM model with $N = 100, ~t_w = 10^5, ~t_m = 10^6$. 
}
\label{fig-3}
\end{figure}

\section{Results and discussions}

In most of the food webs recorded so far only the sign (but not the 
magnitude) of the inter-species interactions are noted. Moreover,  
all the qualitative features of the SM model remain unchanged if, 
instead of allowing the elements $J_{ij}$ to take any real value 
in the domain $[-1,1]$ with equal probability, we allow $J_{ij}$ 
to be only $+1$ or $-1$ with equal probability. Therefore, we work 
with the simpler version of the SM model where each of the elements 
of ${\bf J}$ are discrete binary variables that can be either $+1$ 
or $-1$.  

In our simulations we let the system ``warm up'' for $t_w$ time 
steps, starting from a random initial condition, and then run it 
for further $t_m$ time steps during which collect the data.

\subsection{Distribution of lifetimes} 

It has been claimed in the literature \cite{drossel,newman} that 
the distributions of the lifetimes of the extinct species, estimated 
from the fossil data, can be fitted to a power law of the form 
\begin{equation}
P_t(\tau) \sim \tau^{-x} 
\label{eq-lt}
\end{equation}
where several possible values of $x$, ranging from $1$ to $2$ have 
been suggested. However, the log-log plot of our data in fig.\ref{fig-1}, 
obtained from the simulation of the SM model, demonstrates that an  
approximate power-law-like behaviour is seen only over nearly one 
decade of short lifetimes while strong deviations from power-law  
are seen for longer lifetimes. While the sharp drop of the curves 
in the {\it tail} region can be attributed to {\it finite time} 
effects, the deviations from the early power-law-like region at 
around $\tau \simeq 10$ is neither an artefact of finite time of 
simulation nor that of finite size of the system \cite{chow}.

\subsection{Fitness} 

It was originally pointed out by SM \cite{sole} that 
$f_i = \sum_j J_{ij} S_j - \theta_i$ may be taken as a measure of 
the {\it ``fitness''} of the species $i$; the larger the $f_i$ the 
higher is the fitness of the corresponding species $i$. This 
definition of ``fitness'' is, of course, consistent with the 
expectation in the Darwinian scenario that, at any arbitrary stage of 
evolution, the species with {\it positive} fitness would not become 
extinct. Instead of studying the statistics of $P(f_i)$, or its 
time-average, SM investigated the relation between the species-averaged 
value of $f_i$, i.e., $f_T = \sum_i f_i/N$, and the size of 
the extinctions.

We have now computed the time-averaged probability distribution 
$<P(f_i)>_t$ (following SM, we also take $\theta_i = 0$ for all $i$). 
From this distribution (see fig.\ref{fig-2}) it is clear that, {\it 
on the average}, there are very few species which do not possess 
adequate fitness required for survival and, therefore, become extinct. 
However, the distribution $P(f_i)$ fluctuates with time and exhibits 
significant negative fitnesses when extinction occurs. In these 
stochastic network models of ecosystems, the species become extinct 
because of ``bad luck'' rather than ``bad genes'' \cite{raup}.

We define the incremental change of fitness of a species, say, $i$, 
per unit time step as 
\begin{equation}
\Delta f_i(t) = f_i(t) - f_i(t-1)
\end{equation}
The distribution of $\Delta f_i$ is shown in fig.\ref{fig-3}; it 
falls roughly exponentially in the tail region.

\begin{figure}[tb]
\begin{center}
\includegraphics[angle=-90,width=0.65\columnwidth]{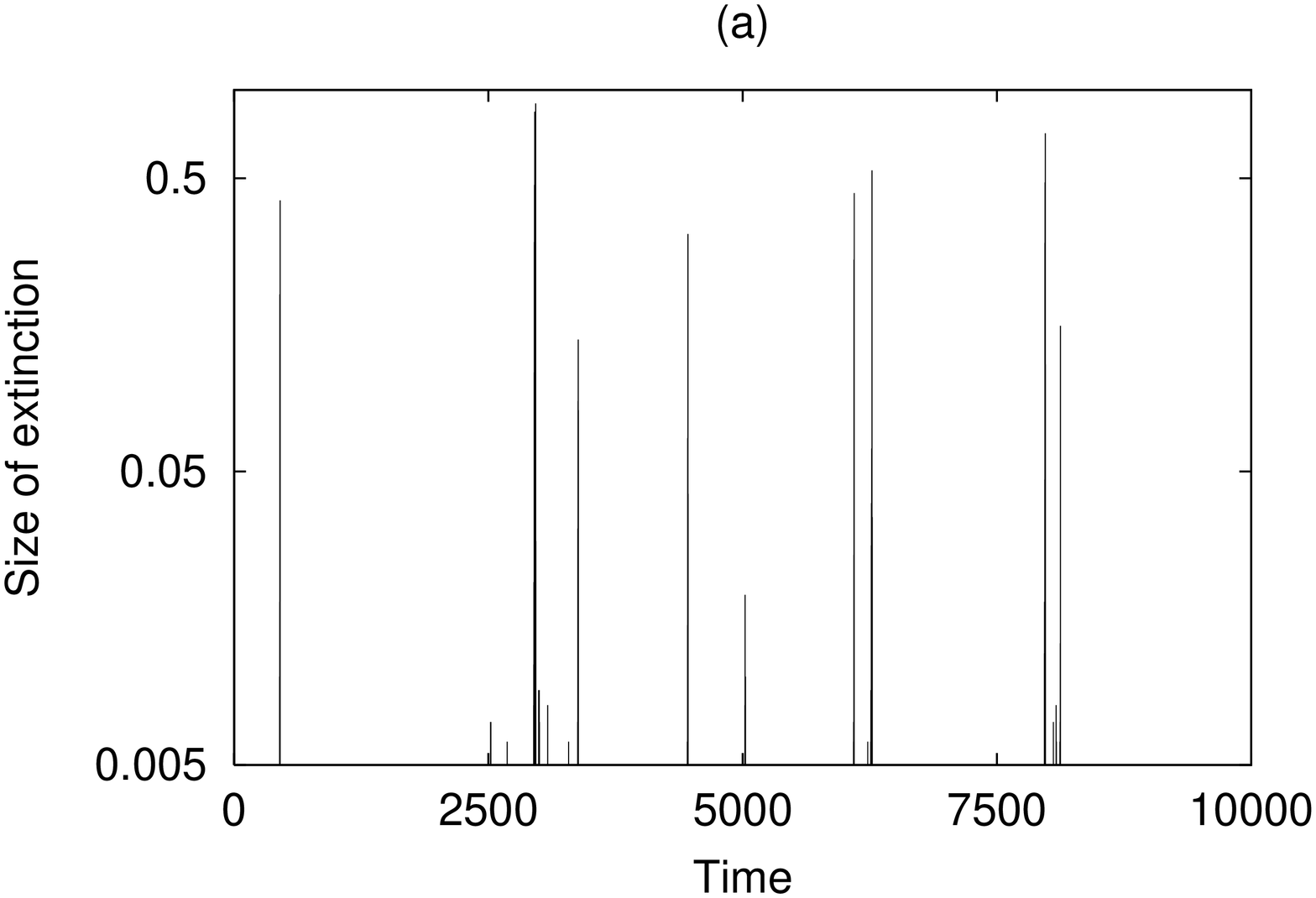}
\includegraphics[angle=-90,width=0.65\columnwidth]{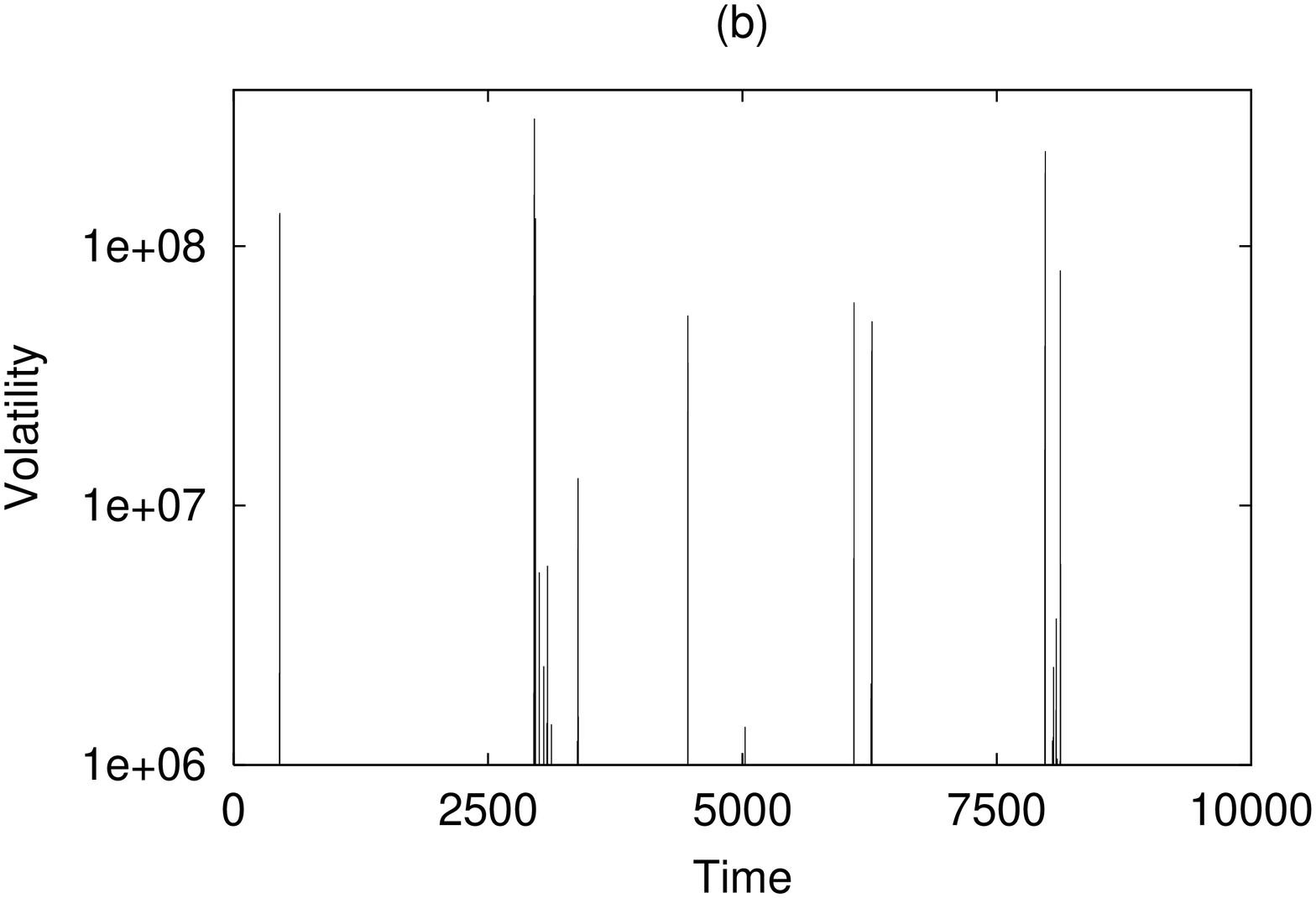}
\end{center}
\caption{The sizes of the extinctions and the volatility are plotted 
against time in (a) and (b), respectively. Data are shown only over 
approximately two and half orders of magnitude, from the top, to 
highlight the strong correlation between the two quantities 
plotted in (a) and (b).}
\label{fig-4}
\end{figure}

\begin{figure}[tb]
\begin{center}
\includegraphics[angle=-90,width=0.65\columnwidth]{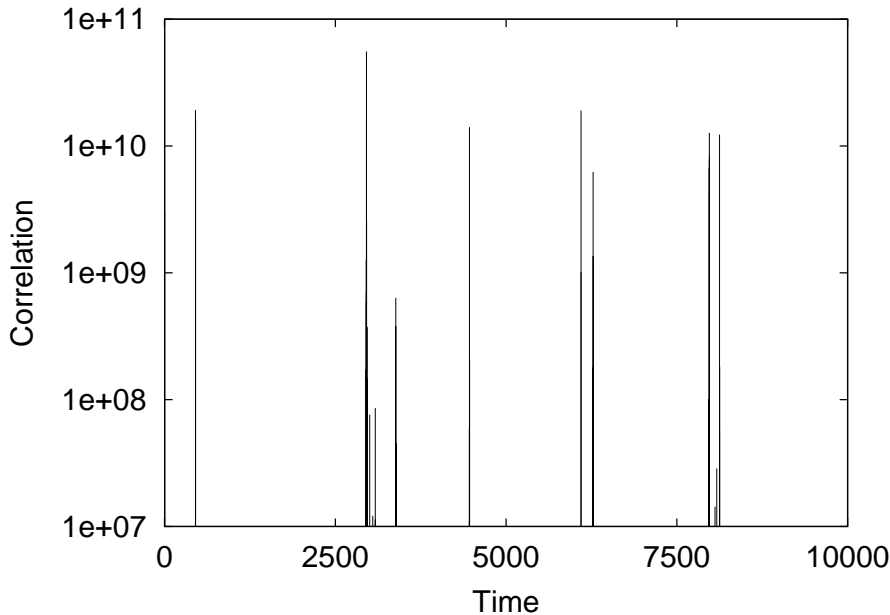}
\end{center}
\caption{The correlation function $C$ is plotted against time. The data 
for $C < 10^7$ are not shown to make the plot less crowded and making 
the comparison with fig.\ref{fig-4} more convenient.}
\label{fig-5}
\end{figure}

\subsection{Volatility and Coherence} 

Borrowing the concept of ``volatility'' in financial markets 
\cite{stanley,bouchaud}, we now define the ``{\it volatility}'' in 
the eco-system by 
\begin{equation}
V(t) = C_1 \sum_i (\Delta f_i)^2
\end{equation}
and the time-dependent correlation function by
\begin{equation}
C(t) = C_2 \sum_{i < j}  (\Delta f_i)(\Delta f_j)
\end{equation}
where the normalization factors $C_1$ and $C_2$ can be chosen 
appropriately. For simplicity, we worked with $C_1 = 1 = C_2$. 
We have also measured the sizes of the extinctions by the 
fractions $m/N$ where $m$ is the numbers of species that 
become extinct simultaneously at the same time step and $N$ 
is the total number of species in the system.

Large extinctions occur when volatility is also large (see 
fig.\ref{fig-4}). $C$ is a measure of ``{\it coherence}'' in the 
changes in the fitness of the species during the same period of 
time. Interestingly, this correlation function is very rarely 
negative. Even when it takes negative value, the corresponding 
magnitude is not large. So, most of the species do not profit 
from the extinction of others.

\subsection{Single common ancestor}

Each node $i$ ($i=1,2,...,N$) carries a label $L(i)$ that indicates 
the identity of its ancestor. In the initial state we assign $L(i)=i$.
During the evolution of the system, whenever a node $\beta$ 
becomes extinct and is, therefore, re-populated by the species 
occupying the node $\alpha$ we re-label the node $\beta$ with 
$L(\alpha)$, i.e., the node $\beta$ also carries the label $L(\alpha)$. 
From our simulation we have always found that, after sufficiently 
long time, all the nodes carry the same label thereby indicating
that all the living organisms at that stage share a single common 
ancestor from which all the wide variety of species have appeared 
as a result of massive diversification. This is perfectly consistent 
with theory \cite{harris} and the fossil record \cite{benton}. The 
distributions of the sizes of phylogenic trees had already been studied 
by Sole and Manrubia \cite{sole}. 

\section{Summary} 

We have computed the distribution of the lifetimes of the species in 
SM model; the deviations from the pure power-law behavior observed 
for the SM model is similar to that in a recent model that describes 
both ``macro''-evolution as well as ``micro''-evolution \cite{chowstau}. 
We have also interpreted $f_i = \sum_j J_{ij} S_j$ as the fitness of 
the species $i$ and investigated the distributions of fitness as well 
as the incremental changes of fitness per unit time. Finally, we have 
introduced the concepts of {\it volatility} and {\it coherence}; 
computing these for the SM model as functions of time, we have 
demonstrated their close relation with the sizes of extinctions.

\section{Acknowledgement} This work is supported in part by the 
Alexander von Humboldt Foundation and by the Deutche Forschungsgeminschaft 
(DFG) through a joint Indo-German research project. We also thank 
the Supercomputer center, J\"ulich, for computer time on their 
CRAY-T3E.

\bibliographystyle{plain}

\end{document}